\documentclass[pra,epsfigure,twocolumn,showpacs]{revtex4}
\usepackage{amsmath}
\usepackage{amstext}
\usepackage{latexsym}
\usepackage{epsfig}
\usepackage{graphicx}
\usepackage{amsfonts}
\usepackage{bm}
\usepackage{amssymb,subfigure}

\newcommand{\ket}[1]{\left\vert#1\right\rangle}
\newcommand{\modul}[1]{\left\vert#1\right\vert}

\newcommand{\bra}[1]{\left\langle#1\right\vert}

\newcommand{\be}{\begin{equation}}
\newcommand{\ee}{\end{equation}}
\newcommand{\bea}{\begin{eqnarray}}
\newcommand{\eea}{\end{eqnarray}}

\newcommand{\N}{{\cal N}}
\newcommand{\eq}[1]{Eq.~(\ref{#1})}

\begin{document}

\title{Enhanced dynamical entanglement transfer with multiple qubits}

\author{A. Serafini,$^{1*}$ M. Paternostro,$^{2*}$ M. S. Kim,$^2$ and S. Bose$^1$}
\affiliation{$^1$ Department of Physics $\&$ Astronomy, University College London,
Gower Street, London WC1E 6BT, United Kingdom\\
$^2$ School of Mathematics and Physics, Queen's University, Belfast BT7 1NN, United Kingdom}
\date{\today}

\begin{abstract}
We present two strategies to enhance the dynamical entanglement transfer from continuous variable (CV) to finite dimensional systems by employing multiple qubits.
First, we consider the entanglement transfer to a composite finite
dimensional system of many qubits simultaneously interacting with a bipartite CV field.
We show that, considering realistic conditions in the generation of CV entanglement, 
a small number of qubits resonantly coupled to the CV system 
is sufficient for an almost complete dynamical transfer of the entanglement. Our analysis also sheds further light on the transition between microscopic and macroscopic behaviours of composite finite dimensional systems coupled
to bosonic fields (like atomic clouds interacting with light). 
Furthermore, we present a protocol based on sequential interactions of the CV system with some
ancillary qubit systems and on subsequent measurements, allowing to probabilistically convert CV entanglement into 
`almost perfect' Bell pairs of two qubits. 
Our proposals are suited for realizations in various experimental settings, ranging from cavity-QED to cavity-integrated superconducting devices.
\end{abstract}

\maketitle
%%%%%%%%%%%%%%%%%%%%%%%%%%%%%%%INTRODUCTION%%%%%%%%%%%%%%%%%%%%%%%%%%%%%%%%%%%%
\section{Introduction}
Due to the infinite dimension of their Hilbert space, continuous variable (CV)
systems allow, in principle, to store an infinite amount of entanglement. 
Even in realistic settings, the entanglement of two-mode squeezed states 
generated in parametric amplification processes with present technology can
be much more than one ebit,
thereby largely exceeding the maximum amount of entanglement between a pair of qubits.
Nonetheless, discrete variable systems -- and, most notably, qubits -- are
naturally privileged for the implementation of many quantum information tasks, 
with the prominent example of quantum computation.
In this respect, it is of interest to investigate the efficiency at which 
entanglement can be transferred from continuous to discrete variable systems 
(by realistic coherent interactions) and to envisage strategies 
to improve such efficiency. 
This is relevant in view of the relative ease with which highly entangled CV states 
can be currently generated~\cite{generate1,laurat}. With efficient transfer procedures, such entanglement could be
distributed to separate discrete variable systems and employed for general quantum information processing purposes.
 
Internal levels of atoms coherently interacting with bosonic light fields 
are natural candidates as discrete variable receivers  
of CV entanglement. 
In fact, the entanglement transfer between two radiation modes and a pair
of two-level atoms 
through coherent interactions has already been investigated \cite{wonmin,ioMEMS,krauscirac}.
Also, the state transfer between macroscopic atomic clouds and 
light has been theoretically considered recently~\cite{kuzmich00} and important steps have been 
made towards its experimental demonstration~\cite{schori02},
relying on measurements and coherent interactions. 
In the limit of a macroscopic number of polarised atoms, 
the atomic component of such systems behaves as a 
CV system, so that the entanglement transfer relates, actually, 
two CV systems of different nature. 
For a resonant Jaynes-Cummings coupling~\cite{JC}, the interaction reduces to the action of a beam splitter
between the two systems allowing, in principle, perfect state (and thus entanglement) transfer.
However, if the receiving system is really a discrete variable one, like a small number of atoms 
({\it i.e.} a ``microscopic'' cloud) would be, then the transfer is no longer perfect.
Some questions of fundamental and practical interest arise in this instance:
How many atoms are actually needed in order to realise an `essentially perfect' transfer,
thus exhausting the resources of the CV system?  
How does the transition between microscopic and macroscopic  
entanglement transfers behave? 

The purpose of this paper is twofold.
On the one hand, we aim to study the entanglement transfer from two 
light modes to a pair of atomic ensembles made up of a small number of
two-level atoms (or, equivalently, to systems with Hilbert spaces of small dimension $d>2$).
We will show that, for two-mode squeezed states realistically achievable in the labs, all the 
entanglement can be extracted by a few atoms, in an essentially microscopic regime. 
Moreover, we will shed further light on the transition from the finite dimensional regime
to the CV behaviour displayed by macroscopic ensembles, explicitly elucidating the algebraic 
reasons lying behind this transition and the physical conditions allowing 
to treat macroscopic ensembles as CV systems.
On the other hand, we propose a method to increase the entanglement transfer to a pair 
of two-level atoms (qubits), by letting further pairs interact successively 
with the same entangled light field and then by postselecting the local measurements on such pairs. 
We will show that, following this route, the entanglement transferred to the first pair of qubits increases considerably with respect to the strategy pursued in Refs.~\cite{wonmin,ioMEMS}. Remarkably, we demonstrate that such an `extraction' 
procedure, aiming at achieving a Bell state of first two atoms, 
can be made ``arbitrarily perfect'' by repeated iterations of the probabilistic protocol.

Both the schemes we suggest in this work are naturally implemented in the context of cavity-quantum electrodynamics (cavity-QED). In this scenario, the qubits would be embodied by two-level atoms interacting with the field modes of two optical cavities externally fed by the two-mode CV system. The feeding process can be accurately described by the effective mixing of the external CV state with vacuum at a beam splitter (represented by the cavity lossy mirror), as detailed in Refs.~\cite{wonmin,ioMEMS,maurotesi}. The trade-off between a cavity which can be externally fed and a sufficiently long coherence time of any intra-cavity photon can be accomplished by exploiting the non-linear effects of the intra-cavity atomic medium as described in Ref.~\cite{turchette} (see also Ref.~\cite{ioMEMS}). Concerning the simultaneous presence of a multi-qubit system in a cavity, 
we recall that an efficient scheme able to infer the actual number of atoms inside an optical resonator has been recently realised, allowing for the exact preparation of up to $3$ atoms inside a single cavity (with the envisaged possibility of extending this number to $\simeq{10}$), over a time three orders of magnitude shorter than the effective trapping time~\cite{kimble}. 
Because of the generality of our protocols, 
other physical systems may be considered in order to implement them. For instance, the postselction scheme can be suitably realised by considering systems of superconducting qubits integrated in quasi-unidimensional cavity-structures~\cite{schoelkopf}, as it will be more explicitly described in Section~\ref{sequential}. 

This paper is organized as follows. In Section~\ref{multiple}, after having
introduced the system at hand and the concept of entanglement transfer, we analyze in detail the extension of this protocol to multiple qubits simultaneously interacting with the CV modes. In Section~\ref{sequential}, we address the case of sequential interactions of indipendent pairs of qubits with the CV modes, followed by proper postselection events (occurring from the second pair on). Finally, in Section~\ref{finale}, we summarize our results.

%%%%%%%%%%%%%%%%%%%%%%%%%%%%%%%%%%%%%%%%%%%%%%%%%%%%%%
\section{Multiple qubits interacting simultaneously}
\label{multiple}

Throughout the paper, we will consider two ensembles of $N$ two-level systems at two separate sites (we will indifferently refer to them as `atoms' or `qubits') 
interacting with a CV system made up of two modes 
of the radiation field (the {\em distributor}), previously prepared in an entangled 
two-mode squeezed state $\ket{\xi_r}$ reading
\be
\ket{\xi_r} = \frac{1}{\cosh(r)}\sum_{n=0}^{\infty} [\tanh(r)]^n \ket{n,n} \; , \label{2msq}
\ee  
where $\ket{n,n}$ stands for the tensor product of two number states $\ket{n}$ 
in each mode, while $r$ is the two-mode squeezing parameter. The entanglement of this 
state increases with increasing $r$, 
with a logarithmic negativity $E_{\cal N}$ given by $E_{\cal N} = 2r/\ln(2)$
(see the following for the definition of $E_{\cal N}$). 
For simplicity, we consider ensembles of two-level atoms, each interacting with the field 
through a Jaynes-Cummings Hamiltonian with the same coupling strength (``Tavis-Cummings'' model \cite{tavis68}).  
The total interaction Hamiltonian of such a system is given, 
taking into account both sites (labelled by $1$ and $2$), by $H_{int}=H_1+H_2$, with 
\be
H_{k} = 
\sum_{j=1}^{N} g (\sigma_{kj}^{+} a_{k} + \,{\rm h.c.})  \; , \label{int}
\ee
where $\sigma_{kj}$ for $k=1,2$ and $j=1,\ldots,N$ is the ladder operator 
referring to atom $j$ in site $k$ ($2\sigma^{+}_k = \sigma_{x}+i\sigma_{y}$ 
in term of the Pauli matrices $\sigma_{x}$ and $\sigma_{y}$), while 
$a_{k}$ stands for the annihilation operator of the radiation mode in site $k$.
Note that the calculations can be easily generalized to the case of asymmetric interaction
which, however, would not significantly affect our main results.
Furthermore, we will restrict to the case 
of a pure state of the CV system. A mixed state can be easily taken into account,
and would simply result in the shrinking of the peak values of the 
transferred entanglement \cite{ioMEMS}, with no substantial changes 
in the qualitative behaviour of the transfer.

We also assume that all the atoms are initialised in the ground state at the beginning 
of the interaction {\em i.e.}, adopting the spin terminology, that they are perfectly ``polarised''.
Note that this assumption, besides being practically satisfiable by the application of a strong polarising field,
is also reasonable, as it has already been shown that, for $N=1$, such an initial atomic configuration allows 
for an optimal entanglement transfer \cite{wonmin}.
Because of this hypothesis, the global state of the atoms (each of which can be seen as a pseudo-spin $1/2$) will stay within the symmetric $(N+1)$--dimensional
subspace with total spin $j=N/2$ during the evolution. 
The dynamics at each site thus reduces 
to the interaction of one mode of the field with
a spin $j$ initially prepared in the ground state $\ket{j,m=-j}$ ($m$ being the value of the projection of the total angular momentum vector along the quantization axis), with raising operator $\sigma^{+}_k$.
The action of $\sigma_k^{+}$ on an atomic state is determined by the Clebsch-Gordan coefficients 
according to $\sigma_k^{+} = \ket{j,m}_k= \sqrt{(j-m)(j+m+1)}\ket{j,m+1}_k$ for $m=-j,\ldots,j$ 
($k=1,2$ referring to the site). 
For convenience, let us relabel the atomic states $\{\ket{j,m}_{k}\}$ at each site
by their number $n$ of `excitations' over the ground state:
$\ket{n}_{k}\equiv\ket{j,n-j}_{k}$ for $n=0,\ldots,N$.

The dynamics at each site can be analytically treated for $N\le 4$~\cite{japan}. 
For instance, let us recall the time evolution operator for $N=1$: 
\begin{equation}
\label{JCrisolto}
{\rm e}^{-i{H}_{k}t}=
\begin{pmatrix}
\cos(\tau\sqrt{a_{k}^{\dag}a_{k}+1})&-ia_{k}\frac{\sin(\tau\sqrt{a_{k}^{\dag}a_{k}})}{\sqrt{a_{k}^{\dag}a_{k}}}\\
-ia_{k}^{\dag}\frac{\sin(\tau\sqrt{a_{k}^{\dag}a_{k}+1})}{\sqrt{a_{k}^{\dag}a_{k}+1}}&\cos(\tau\sqrt{a_{k}^{\dag}a_{k}})
\end{pmatrix},
\end{equation}
where the matrix acts on the ordered atomic two-dimensional Hilbert space $\{\ket{1},\ket{0}\}_k$ (whereas $a$ and $a^{\dag}$ act
on the infinite-dimensional Hilbert space of the field), $\tau\equiv gt$ is the rescaled interaction time and $\hbar=1$.  
We have numerically integrated the dynamics for larger number of atoms (up to $N=20$),
truncating the infinite dimensional Hilbert space of the field modes at number states 
depending on the initial squeezing parameter $r$ (obviously, as $r$ increases higher excitation
states get more significantly populated). 
For the sake of clarity, let us explicitly describe the dynamics of the system under the action of the
evolution operator $\exp(-iH_{int}t)=\exp(-iH_{1}t)\otimes\exp(-iH_{2}t)\equiv U_1\otimes U_2$. The density matrix $\varrho(\tau)$ describing the state of the system at 
the rescaled time $\tau$ reads
\be
\varrho(\tau) = U_1\otimes U_2 (\ket{\xi_r}\boxtimes \ket{0,0})(\bra{0,0}\boxtimes \bra{\xi_r})
U_1^{\dag}\otimes U_2^{\dag} \, , \label{density}
\ee
where $\otimes$ denotes the tensor product with respect to the 
bipartition in the two sites, $\boxtimes$ denotes the tensor product with respect to 
the bipartition into the atomic and radiation subsystems and the state 
$\ket{0,0}=\ket{0}_1\otimes\ket{0}_2$ stands 
for the ground state of the global atomic system.
We are now interested in the reduced atomic state $\varrho_{at}(\tau)$, obtained by partially tracing the global state $\varrho(\tau)$ over the field variables:
$
\varrho_{at}(\tau) = \,{\rm Tr}_f \varrho(\tau)
$, 
${\rm Tr}_f$ denoting the trace over the Hilbert space of the field.

\begin{table}[b!]
\begin{center}
\begin{tabular}{|c|c|c|c|}
\hline
\hspace*{2mm}$N$\hspace*{2mm} & \hspace*{2mm}$r_{opt}$\hspace*{2mm} & \hspace*{2mm}$E_{\N}$\hspace*{2mm}
&\hspace*{2mm}$eff$\hspace*{2mm}\\
\hline \hline
$1$ & $0.86$ & $0.90$ & $0.36$ \\
\hline 
$2$ & $0.86$ & $1.25$ & $0.50$ \\
\hline 
$3$ & $0.92$ & $1.59$ & $0.60$ \\
\hline 
$4$ & $1.02$ & $1.85$ & $0.63$ \\
\hline 
$5$ & $1.10$ & $2.07$ & $0.65$ \\
\hline 
\end{tabular}
\hspace*{0.2cm}
\begin{tabular}{|c|c|c|c|}
\hline
\hspace*{2mm}$N$\hspace*{2mm} & \hspace*{2mm}$r_{opt}$\hspace*{2mm} & \hspace*{2mm}$E_{\N}$\hspace*{2mm}
&\hspace*{2mm}$eff$\hspace*{2mm}\\
\hline \hline
$6$ & $1.18$ & $2.24$ & $0.66$ \\
\hline
$7$ & $1.22$ & $2.40$ & $0.68$ \\
\hline 
$8$ & $1.27$ & $2.54$ & $0.69$ \\
\hline 
$9$ & $1.31$ & $2.66$ & $0.70$ \\
\hline 
$10$ & $1.35$ & $2.77$ & $0.71$ \\
\hline 
\end{tabular}
\end{center}
\caption{Optimal values of the two-mode squeezing parameter, maximal 
logarithmic negativity achievable (through resonant interaction with a CV state with the optimal 
two-mode squeezing) and `efficiency' $eff$ of the optimal transfer [quantified by the ratio 
between the transferred logarithmic negativity and the logarithmic negativity of the 
initial distributor's CV state: $eff\equiv E_{\N}\ln(2)/(2r)$]
for the first values of $N$. Note the monotonic increase of the ratio $eff$. \label{table}}
\end{table}

Finally, the `logarithmic negativity' $E_{\cal N}$ of the state $\varrho_{at}(\tau)$ can be worked out 
to estimate the rate of entanglement transferred from the CV two-mode squeezed state to 
the finite dimensional Hilbert space of the atoms. 
Let us recall that such a quantity  
is strictly related to the positivity of the partial transpose (PPT) criterion
for the separability of quantum states \cite{npt}.
The partial transposed density matrix $\tilde\varrho$ is obtained from any given bipartite quantum 
state $\varrho$ by transposing the variables of only one of the two subsystems. The PPT criterion then 
simply reads $\tilde{\varrho}\ge0$.
Indeed, such a criterion is necessary for the separability of any quantum state regardless of the 
dimension of the system's Hilbert space but is also sufficient only for $2\times 2$ 
and $3\times 2$ Hilbert spaces (considering only finite-dimensional cases).
\begin{figure}[t!]
\begin{center}
\includegraphics[scale=0.6]{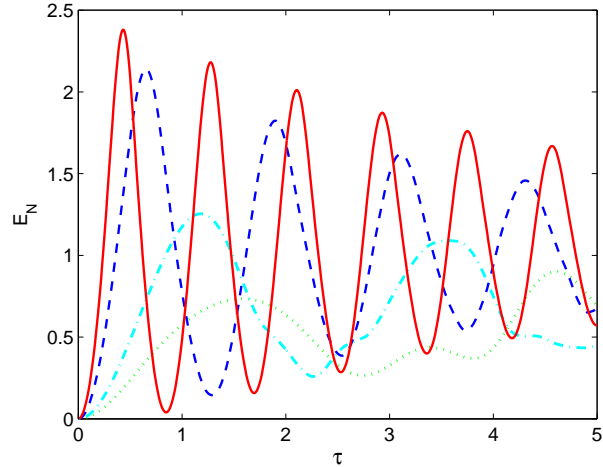}
\caption{(color online). Entanglement transferred from a two-mode squeezed state with $r\simeq 0.86$ 
(with initial $E_{\N}\simeq 2.48$) for various number $N$ of receiving two--level atoms at each site.. 
The red curve refers to $N=15$ (solid line), the blue curve to $N=7$ (dashed line), the cyan curve to $N=2$ (dash-dotted line) and 
the green curve to $N=1$ (dotted line). All quantities plotted are dimensionless.\label{interfacemany}}
\end{center}
\end{figure} 
\begin{figure*}[t!]
\centering
 \subfigure[\label{transfer3}]
{\includegraphics[width=4.1cm]{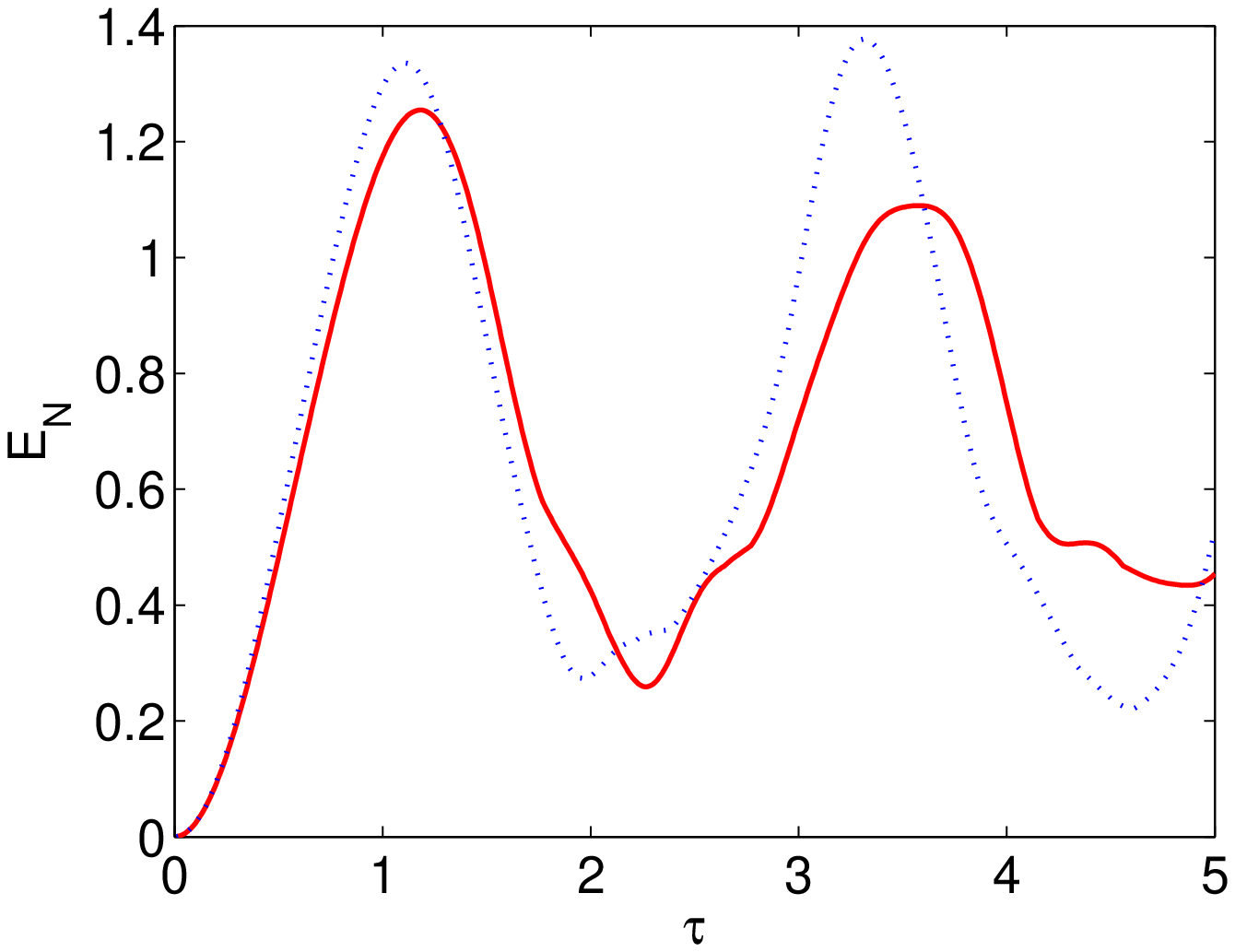}}%6.7
 \hspace{4mm}
\subfigure[\label{transfer8}]
{\includegraphics[width=4.1cm]{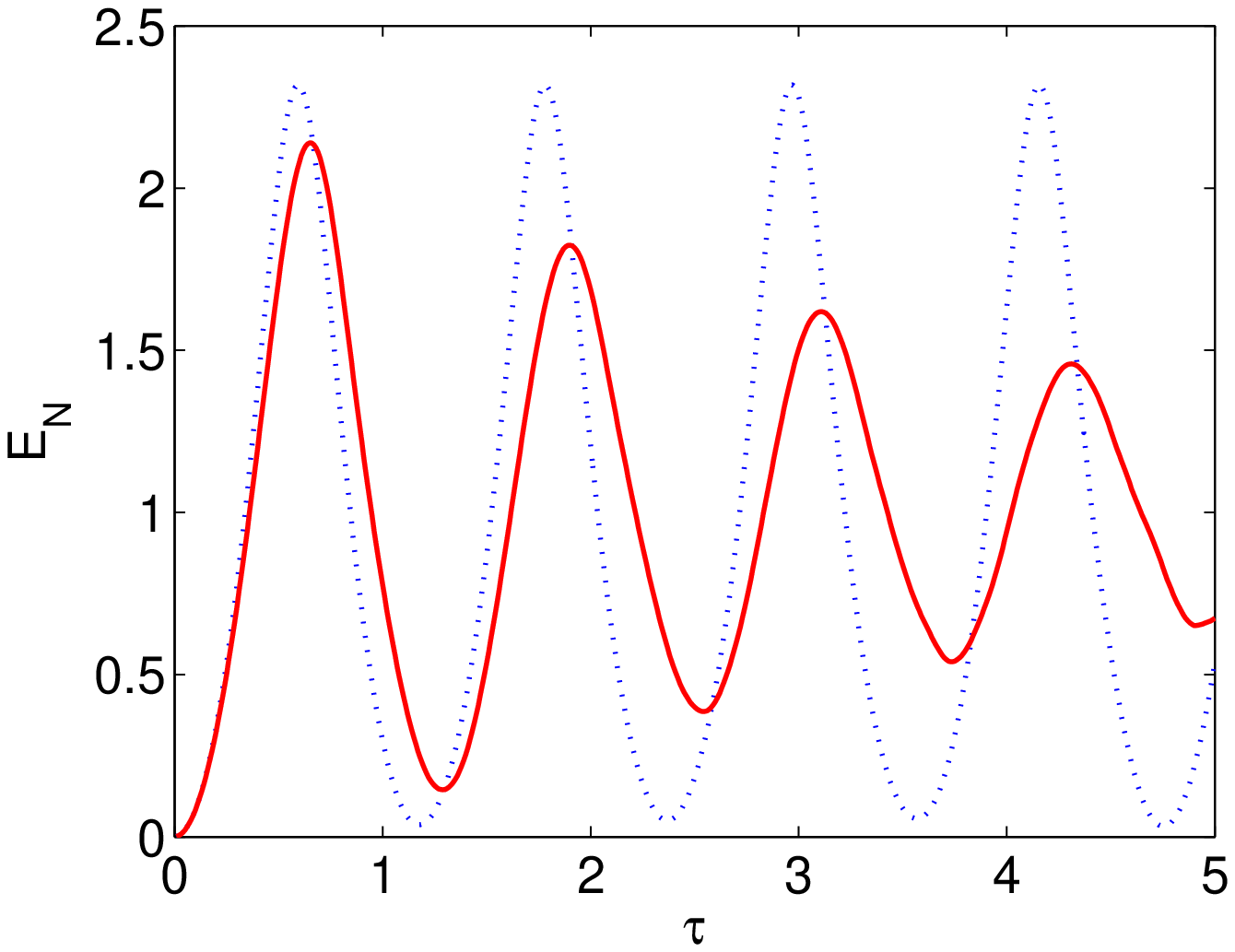}}%6.9
 \hspace{4mm}
\subfigure[\label{transfer12}]
{\includegraphics[width=4.1cm]{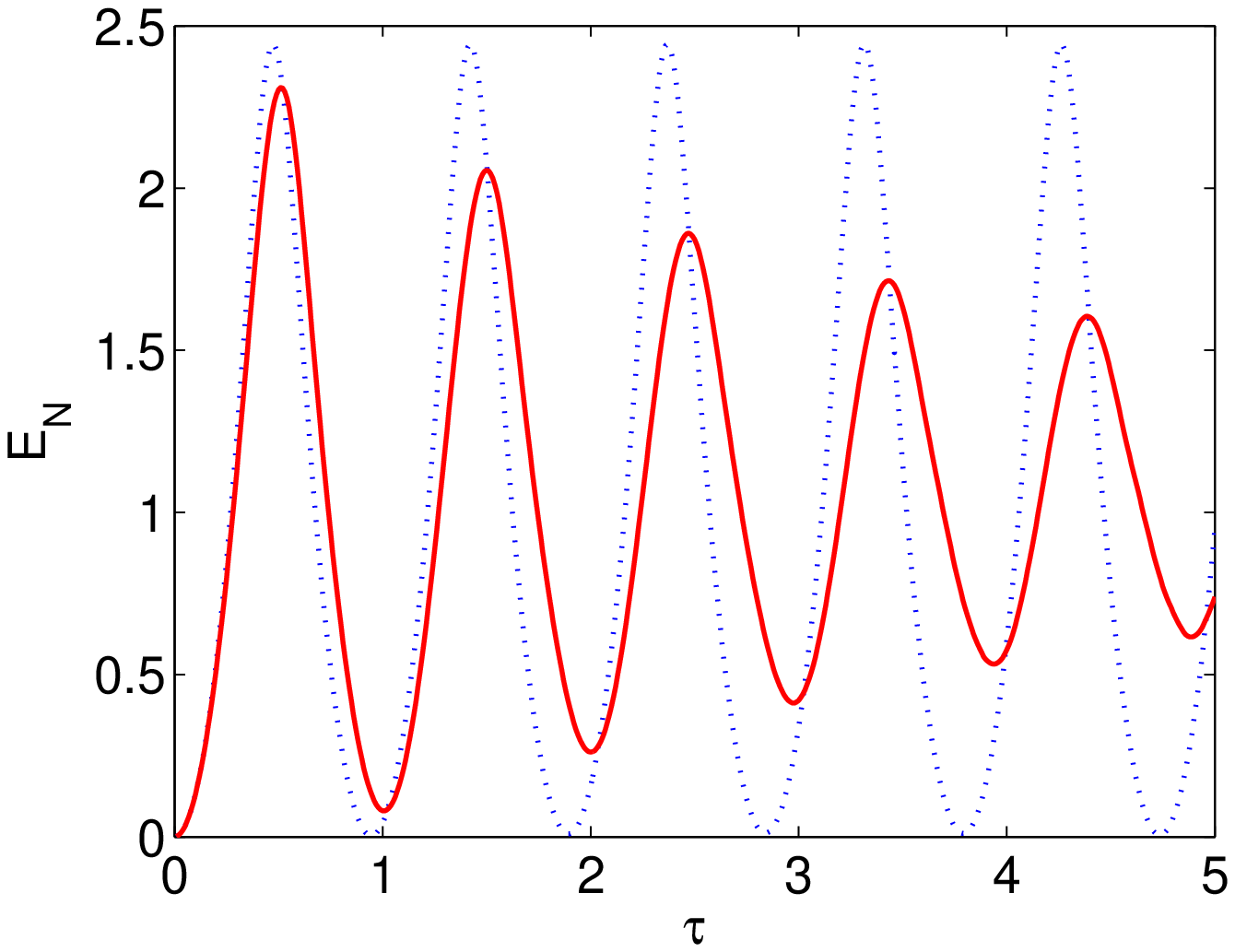}}%7.0
 \hspace{4mm}
\subfigure[\label{transfer16}]
{\includegraphics[width=4.1cm]{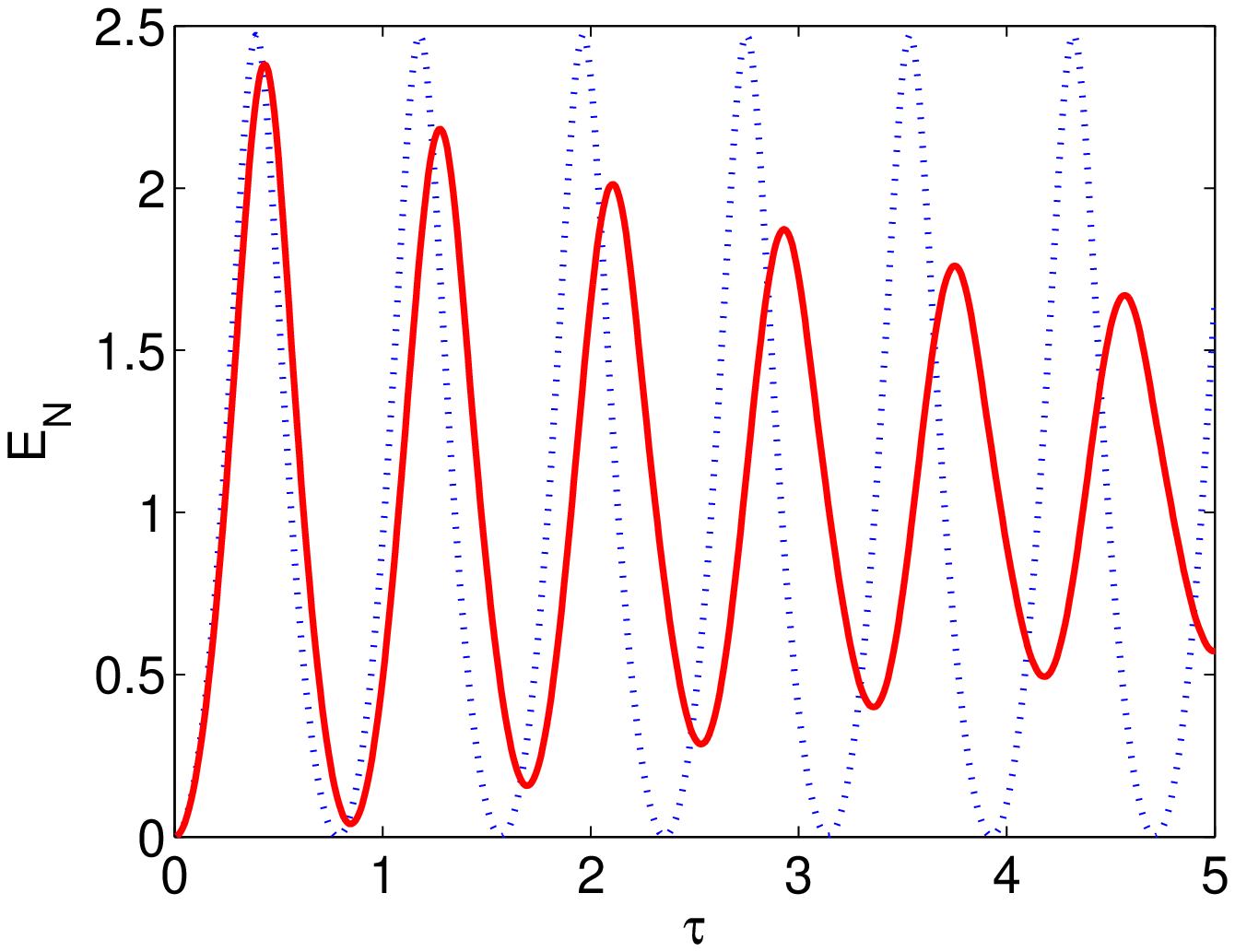}}%7.1
  \caption{(color online). Comparison between the entanglement transfer to pair of ensembles of $N$ atoms (solid lines) 
   and to pairs bosonic modes with interaction strength rescaled by $\sqrt{N}$
   (dotted lines), for N=2 (a), for N=7 (b)
   , for N=11 (c), for N=15 (d). All quantities plotted are dimensionless.}
\label{transfer}
\end{figure*}
The `negativity' $\N$ (first envisaged in Refs.~\cite{zircone}, later
thoroughly discussed in Refs.~\cite{jenstesi}), which can be easily determined from the knowledge of the density matrix (upon diagonalisation of $\tilde{\varrho}$), is simply defined as the absolute value of the sum 
of the negative eigenvalues of $\tilde\varrho$ and directly quantifies the violation of the PPT criterion.
The logarithmic negativity is then just given by $E_{\N}\equiv \log_{2}(2\N+1)$ (so that $E_{\N}=1$ for Bell states of two qubits).  
From an operational point of view, the quantity $E_{\N}$ is an
upper bound to the {\em distillable entanglement} 
and is related to the entanglement cost under PPT
preserving operations \cite{auden03}.
Let us mention that, since the PPT criterion is no longer sufficient for Hilbert spaces of dimension
$3\times 3$ and higher, such a quantity is null for some entangled states in higher dimensions.  
Still, $E_{\cal N}$ has been recently proven to be an entanglement monotone 
irrespective of the dimension of the Hilbert space under study \cite{plenio05}. Moreover, it satisfies the other 
crucial requirement of being null for non entangled states. Thus, it can be employed as a proper quantification
of the entanglement for high-dimensional Hilbert spaces. 
Clearly, increasing the dimension of the Hilbert space allows for ``more space'' being available for entanglement: in fact, the maximal value of $E_{\N}$ is given by $\log_{2}(d)$ for a bipartite $(d\times d)-$dimensional Hilbert space.

Let us now move on to analyse the entanglement transfer to pairs of ``microscopic'' ensembles of
$N$ atoms.
Interestingly, as already found in the case $N=1$ \cite{wonmin}, the maximal entanglement 
which can be dynamically transferred to a finite dimensional system is not monotonically increasing 
with the entanglement of the initial CV state ({\em i.e.}~with $r$).
Indeed, the value of $r$ allowing for the optimal transfer (after maximisation over time), 
which we will denote by $r_{opt}(N)$, increases with increasing $N$ but is always finite. 
The behaviour of the
function $r_{opt}(N)$, whose values for small $N$ are reported in Table~\ref{table},
is not trivial, 
resulting from a compromise between the amount of entanglement initially 
present in the system (monotonically increasing with $r$) and a population of the excitated states of the field modes favouring a better state transfer (too large $r$'s imply large
populations in higher number states and, as a consequence, a mismatch of the `effective' 
dimensions of atoms' and light's Hilbert spaces). 
Notice that  
$$\lim_{N\rightarrow\infty} r_{opt}(N) = \infty \; , $$ 
as in the infinite dimensional (CV) limit a beam-splitter allows for perfect state transfer  
and, in principle, an infinitely entangled initial state (obtained in the limit $r\rightarrow\infty$)
can be also perfectly transferred to the atoms. 

Also, one has $r_{opt}(1)\simeq 0.86$, for which 
the initial CV state $\ket{\xi_{r_{opt}(1)}}$ 
has a logarithmic negativity $\bar E_{\N}\simeq 2.48$. 
For such a two-mode squeezing parameter, which is in the range of current experimental 
techniques \cite{generate1,laurat}, we have investigated the possibility of realising 
`exhaustive' entanglement transfer from
CV to discrete variable systems by increasing the number of atoms $N$. 
Fig.~\ref{interfacemany} shows the behavior of the transfer for different values of $N$. 
The efficiency and velocity of the transfer increase with increasing $N$, the value 
of 15 allowing for a very good transfer to the 16--dimensional atomic space
(with $E_{\N}\simeq2.38\simeq0.96\bar E_{\N}$). Let us mention that the truncation 
of the CV state to the state $\ket{15,15}$ has a logarithmic negativity $E_{\N}\simeq2.47$,
thus containing almost all the entanglement of the CV state.
In general, for the squeezing parameters currently achievable, 
a small number of atoms ($N\simeq10-100$) is always capable to retrieve  
a substantial part of the entanglement contained in the initial two-mode squeezed state.
For instance, the most entangled two-mode state reported in Ref.~\cite{laurat}
has a logarithmic negativity $E_{\N}\simeq1.60$, corresponding to $r\simeq{0.55}$~\cite{proviso}.
For such a value of $r$, a number of atoms $N=10$ allows for the transfer
of a logarithmic negativity $E_{\N}\simeq 1.55$, whereas for $N=20$ one reaches
the value $E_{\N}\simeq 1.58$, thus almost exhausting the resources of the
initial CV state.
Indeed, values of $N$ of this order allows for an almost perfect extraction
of the entanglement even in instances still far from the current experimental
possibilities. To be more specific, let us consider the most entangled state
considered in the first of Ref.~\cite{laurat}, with a logarithmic negativity
$E_{\N}\simeq 4.53$ corresponding to $r\simeq1.57$.
Now, the logarithmic negativity of this CV state is mostly contained into the first 
$40$ levels of the field's Hilbert space (the truncation of the CV state to the state $\ket{40,40}$ has $E_{\N}\simeq 4.49$).
Even though a direct numerical treatment of this instance would be cumbersome, one can provide 
a very good estimate of the logarithmic negativity which can be transferred to $40$ atoms 
by assuming that the same ratio between the maximal transferable logarithmic negativity and the 
initial available one as in the case $r=0.86$ and $N=15$ applies here.
Note that this a very conservative assumption, as the efficiency of the transfer generally increases 
with increasing $N$ since, as we will appreciate in detail shortly, the finite dimensional system better mimic a 
quantum harmonic oscillator (see also Table \ref{table}). 
Finally, we infer that a logarithmic negativity $E_{\N}\gtrsim 4.31$ 
could be transferred from the CV to the finite 
dimensional system, by letting interact $N\simeq40$ qubits simultaneously.

As we have seen,
while being a great improvement with respect to the transfer to single atoms, 
the ``microscopic'' transfer is never completely exhaustive, even though the receiving 
Hilbert space would be able to almost perfectly contain the initial CV state. 
We now show that the reason for this fact lies in the dynamics of the system, 
and not in the mismatch between the different Hilbert spaces.
To this aim, 
let us consider in more detail the features of the transition of the atomic system 
between the microscopic and the macroscopic regime which, as well known, can mimic 
under certain conditions a CV system interacting through a beam-splitter.
Firstly, let us remark that, if the ladder operators of the atomic ensembles $\sigma^{+}_{k}$
were behaving like actual bosonic ladder operators $b^{\dag}_k$, {\em i.e.}~according to 
\be
b^{\dag}_{k} \ket{n}_k= \sqrt{n+1} \ket{n+1}_{k}
\ee
(with the additional proviso $b^{\dag}_k \ket{N}_k = 0$),
the state (and thus entanglement) transfer from light to the atoms would be essentially perfect 
for low enough squeezing parameters (as the only requirement on the atomic subsystem would 
consist in having enough dimensions to contain the initial field's state up to its effective truncation),
with perfect (not `convoluted') oscillations resulting from a beam-splitter-like dynamics.
On the other hand, 
the action of the actual operators $\sigma^{+}_k$ is described by 
\be
\sigma^{+}_{k} \ket{n}_k= \sqrt{(N-n)(n+1)} \ket{n+1}_{k} \; 
\ee
for $n=0,\ldots,N$.
Let us now focus on the subspaces $\Sigma_{k}$ (for k=1,2) of the atomic Hilbert spaces, 
spanned by basis-vectors $\ket{n}_k$ such that $N\gg n$ (corresponding to the macroscopic, 
``highly polarised'' limit). Denoting by $\sigma^+_{\Sigma_k}$ and $b_{\Sigma_k}$ 
the restrictions of $\sigma^+_{k}$ and $b_{k}$ to the 
subspace $\Sigma_{k}$ one has, 
as apparent from the previous two equations, $\sigma^+_{\Sigma_k}\simeq \sqrt{N} b^{\dag}_k$,
which properly accounts for the bosonic-like behaviour of the polarised atomic system.
Notice that this relationship also implies that,
on the `highly polarised' subspace, the SU(2) algebra approximates 
the Heisenberg algebra of the canonical commutation relations, 
because the restriction of the operator $\sigma^z$ 
can be regarded as a (large) constant: $\sigma^z_{\Sigma_k}\simeq N/2$. 
Explicitly, $[\sigma^-_{\Sigma_k},\sigma^+_{\Sigma_k}]\simeq N\simeq 2\sigma^z_{\Sigma_k}$, 
as one should expect.
Let us note that
the $\sqrt{N}$ factor relating the ladder operators to their bosonic approximations 
accurately explains the 
aforementioned speed-up in the transfer with increasing $N$.  
To explore in detail the transition to the macroscopic regime,
we have numerically compared the (real) entanglement transfer obtained with the atomic 
operators $\sigma^{+}_k$ to the ideal transfer obtained by replacing them with the 
operators $b^{\dag}_{k}$. As shown in Fig.~\ref{transfer}, the mimicking is still not very 
accurate even for $N=15$, thus explaining the good, but not quite perfect entanglement transfer 
in this instance
(even though all the needed capability would be available in the receiving Hilbert space).
Moreover, in all the plots, it is apparent that the matching with the regular bosonic oscillations
is lost with increasing interaction time: this is due to the fact that higher atomic levels
(with $n$ not negligeable with respect to $N$) get 
excited and the atoms lose their original polarization. In fact, the bosonic approximation 
commonly employed in treating the interaction of light with atomic ensembles is not only 
restricted to the macroscopic and polarised case, but also to short interaction times.  

\begin{figure}[t!]
\centering
 \subfigure[\label{fieldmes2}]
{\includegraphics[width=8.8cm]{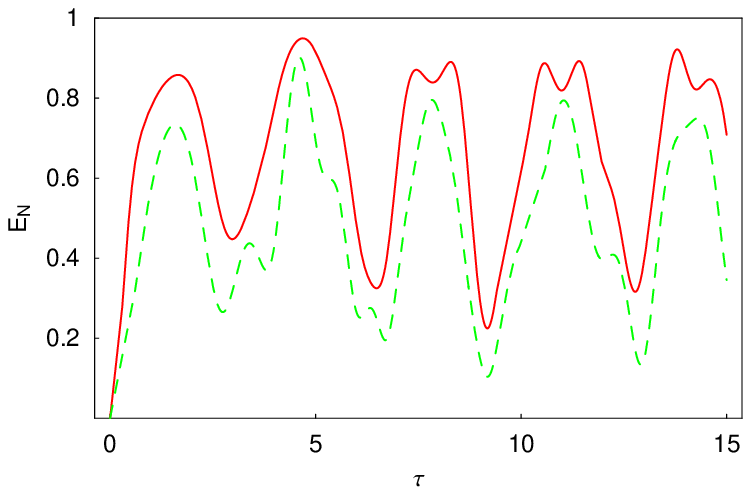}}%6.7
 \hspace{5mm}
\subfigure[\label{fieldmes3}]
{\includegraphics[width=8.4cm]{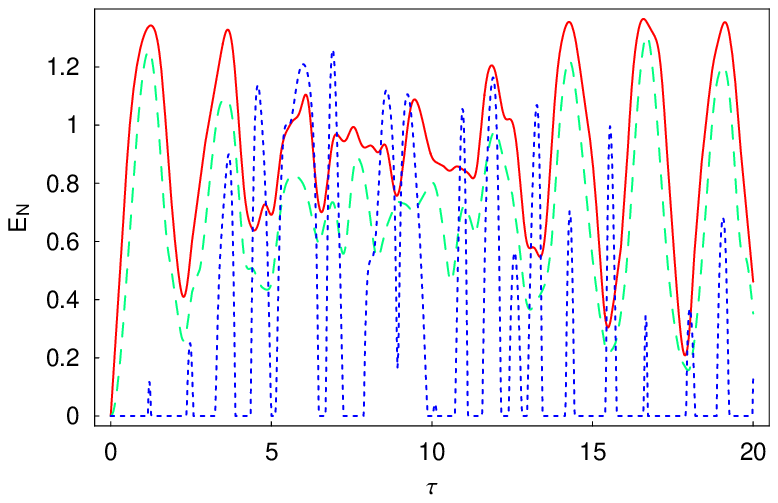}}%6.9
  \caption{(color online). Entanglement transfer for $N=1$ (a) and $N=2$ (b) after postselection
  of the finite dimensional state for an even (solid line) and odd
  [dotted line, always null in case (a)] CV system's photon number. The measurements are taken after the CV and qubit systems have
  interacted for a rescaled time $\tau$. The non postselected,
  `purely dynamical' entanglement transfer is plotted as a reference (dashed
  line). All quantities plotted are dimensionless.}
\label{fieldmes}
\end{figure}

As a final remark about the entanglement transfer to small atomic ensembles, let us mention the possibility, already addressed in
Ref.~\cite{ioNJP}, of probabilistically
improving the performance of the transfer by measurements on the CV system and subsequent postselection. In particular, we have here considered
the measurement of the parity of the field's photon number. The results of
this study for the cases $N=1$ and $N=2$ are reported in Fig.~\ref{fieldmes}.
In general, the outcome ``even'' always increases the entanglement of the finite dimensional state. Conversely, the outcome ``odd'' always completely
spoils the entanglement for $N=1$ (corresponding to the tranfer to two two-level
systems) as, because of the structure of the state resulting from the
dynamical evolution, it effectively projects such a state onto a separable subspace. Still, notice that for $N>1$ the outcome ``odd'' may actually enhance the entanglement, depending on the time at which the measurement is performed.

%%%%%%%%%%%%%%%%%%%%%%%%%%%%%%%%%%%%%%%%%%%%%%%%%%%%%%%%
\section{Multiple atoms interacting successively}
\label{sequential}

As we have seen in the previous section, simultaneous interactions with many atoms allow for 
a very efficient entanglement extraction from infinite to finite dimensional systems. 
Despite this remarkable fact, the main interest of such entangling schemes lies in the possibility 
of achieving highly entangled states of pairs of two-level systems, which are privileged and archetypical 
for computational and communication protocols.  
In the present section we present a strategy (based on the measurement of ancillary qubit
subsystems and on the subsequent postselection) to enhance the entanglement extraction 
from the CV system to two qubits. 
In general, such schemes result in the 
purification of the interesting two-qubit state and thus in a sort of entanglement `purification', 
as is the case for the postselection of detection events of the field modes (see the previous section 
and Ref.~\cite{ioNJP}). 
We mention that a scheme employing ancillary qubits and measurements was also proposed in Ref.~\cite{krauscirac}, addressing the realization of quantum repeaters through atom-light interactions.

\begin{figure}[t!]
\psfig{figure=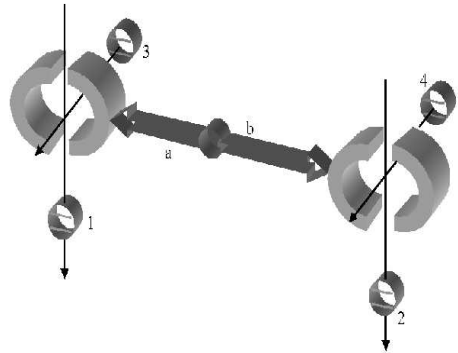,width=6.0cm,height=4.7cm}
\caption{Scheme of the protocol based on sequential passage of a multi-qubit system, shown for just two qubit pairs. Two qubits, labelled $1$ and $2$, interact with the entanglement distributor, represented by a two-mode squeezed state (generated by the non-degenerate-parametric amplifier (NDPA)). Then, a second pair of qubits, $3$ and $4$, are sent to interact with the entangler and are subsequently measured (for instance, by means of two channeltrons) onto the qubit-basis $\{\ket{\psi_{\alpha}},\ket{\psi_{\alpha}}^{\bot}\}$ (defined in the body of the paper).} \label{schema}
\end{figure}

Here, 
instead of measuring the state of the CV subsystem (which might present some practical difficulties), we take in exam the measurement of the states of ancillary qubit-pairs. The scheme of principle of the protocol we consider here is sketched in Fig.~\ref{schema}. In order to elucidate the basics of the protocol, we consider just two pairs of qubits, the first one given by qubits $1$ and $2$ and the second by qubits $3$ and $4$. The first pair interacts with the entanglement distributor for a time $\tau_1$, after which the system is described by the density matrix $\varrho(\tau_1)$ determined by \eq{density} for $N=1$. 
Then, the second pair of qubits interacts with the CV system for an interaction time $\tau_2$. Each qubit interacts with the local light mode according to the Jaynes-Cummings Hamiltonian of Eq.~(\ref{int}).
After such a dynamical evolution, qubits $3$ and $4$ are individually measured in the orthogonal basis 
$\{\ket{\psi_{\alpha}},\ket{\psi_{\alpha}}^{\bot}\}$ with $\ket{\psi_{\alpha}}\equiv
\alpha\ket{0}+\beta\ket{1}$. Here, $|\beta|^{2}=1-|\alpha|^2$ and $|\alpha|\in[0,1]$ (for simplicity, we assume the same 
measurement is performed on the two qubit ancillary systems~\cite{commento1}). 
We consider all the qubits initialized in the ground state $\ket{0}$ and
the entangled state $\ket{\xi_r}$ of \eq{2msq} as the initial state of the field, constituting the entanglement resource. 
Let us recall, once more, that this choice for the initialization of the qubits, other than being natural, has been proven to optimize the entanglement transfer process~\cite{wonmin,ioJJ,maurotesi}. In what follows, we set the value of the two-mode squeezing parameter $r$ to $0.86$ which, as already remarked, allows for the maximal transfer in the ``passive'' scheme with $N=1$.
We denote by $U_3$ and $U_4$ the evolution operators that describe the dynamics of 
qubits $3$ and $4$ interacting with the light modes 
[each of them being given by Eq.~(\ref{JCrisolto}) for $\tau=\tau_2$].
Let us also suppose that the measurement projects qubits 3 and 4 onto the 
two-qubit state $\ket{cd}\equiv\ket{\psi_{\alpha}}\otimes\ket{\psi_{\alpha}}$.
After such a measurement, qubits $1$ and $2$ are left in the state $\varrho_{cd,12}$ reading  
\be
\varrho_{cd,12}=\frac{{\rm Tr}_{34,f}[\ket{cd}\bra{cd}(U_3\otimes U_4) \varrho(\tau_1) 
(U_3^{\dag}\otimes U_4^{\dag})]}{{\rm Tr}\,[\ket{cd}\bra{cd}(U_3\otimes U_4) \varrho(\tau_1) 
(U_3^{\dag}\otimes U_4^{\dag})]} \, ,
\label{densitycd}
\ee
with probability 
\begin{equation}
\label{prob}
p={\rm Tr}\,[\ket{cd}\bra{cd}(U_3\otimes U_4) \varrho(\tau_1) (U_3^{\dag}\otimes U_4^{\dag})].
\end{equation}
Here ${\rm Tr}_{34,f}$ stands for the partial trace on the Hilbert space of qubits $3$ and $4$ and on the field while ${\rm Tr}$ is the usual global
trace.

Before proceeding with the quantitative description of the performances of this protocol, it is worth remarking that the addressed situation can be realized in different physical setups such as cavity-QED systems, individually-trapped 
ions interacting with optical cavities~\cite{kaler} and cavity-integrated superconducting systems~\cite{schoelkopf,ioMEMS,ioJJ}. 
In the first case, the two pairs of qubits could cross an optical cavity along two orthogonal directions on the same plane (see Fig.~\ref{schema}) \cite{commento2}. In the second and third cases, a specific qubit-pair can be set in resonance 
with a cavity-field mode by tuning an external magnetic flux which, at the same time, sets the qubits of the second pair in a dispersive (far off-resonant) regime. The technology needed for this kind of operations already exists, requiring a magnetic field with a spatial gradient. The magnetic field globally addresses the two qubits of two different pairs which are simultaneously present inside a single cavity, but differently Zeeman-shifts them, setting one qubit in resonance and the other off-resonance (in a way which extends and generalizes the Mintert-Wunderlich proposal~\cite{wunder}). However, while the strong-coupling regime seems to be remotely feasible in the cavity-trapped ion setup, the main advantage which characterizes the superconducting implementation is the natural achievement of these working conditions. This is due to the fact that the electromagnetic fields, in a
system of superconducting charge qubits~\cite{ioJJ}, couple directly to the excess charge of the qubit rather than to its intrinsic dipole moment. This gives rise to effective dipole moments which are orders of magnitude larger than the already large ones characterizing alkali Rydberg atoms interacting with cavity-field modes~\cite{schoelkopf,ioMEMS}. Moreover, the very high cavity quality factors achievable in this quasi-unidimensional configuration~\cite{schoelkopf,juzelinas} and a judicious choice of the superconducting qubits working point (the charge-degeneracy point, as described in Refs.~\cite{ioJJ,schoelkopf,ioMEMS}), which strongly quenches the non-Markovian $1/f$ noise~\cite{elisabetta,ioMEMS,maurotesi}, make our proposal robust against decoherence. Thus, a suitable modification of already realizable fully integrated setups seems to be very promisig for the implementation of the present proposal. 

As for the scheme of principle of our proposal, 
note that an alternative approach based on the simultaneous interaction of each light field with 
$N$ qubits (as in the previous section) and on the subsequent measurements of $N-1$ qubit systems 
to {\em localize} the entanglement in the interesting pair might have been considered. 
Actually, even allowing for 
generic qubit measurements, such a strategy cannot beat the direct dynamical transfer to two qubits
(obtained for $N=1$):
the ``dispersion'' of the entanglement among the qubit subsystems cannot be reversed by local 
measurements and classical communication. 
The projection of qubit-pairs which sequentially interact with the distributor, as is the case for the scheme presented here, is thus a key to improve the `passive', purely dynamical transfer.

\begin{figure}[t]
\centerline{{(a)}\hskip4.0cm{(b)}}
%\vskip-2cm
\centerline{\psfig{figure=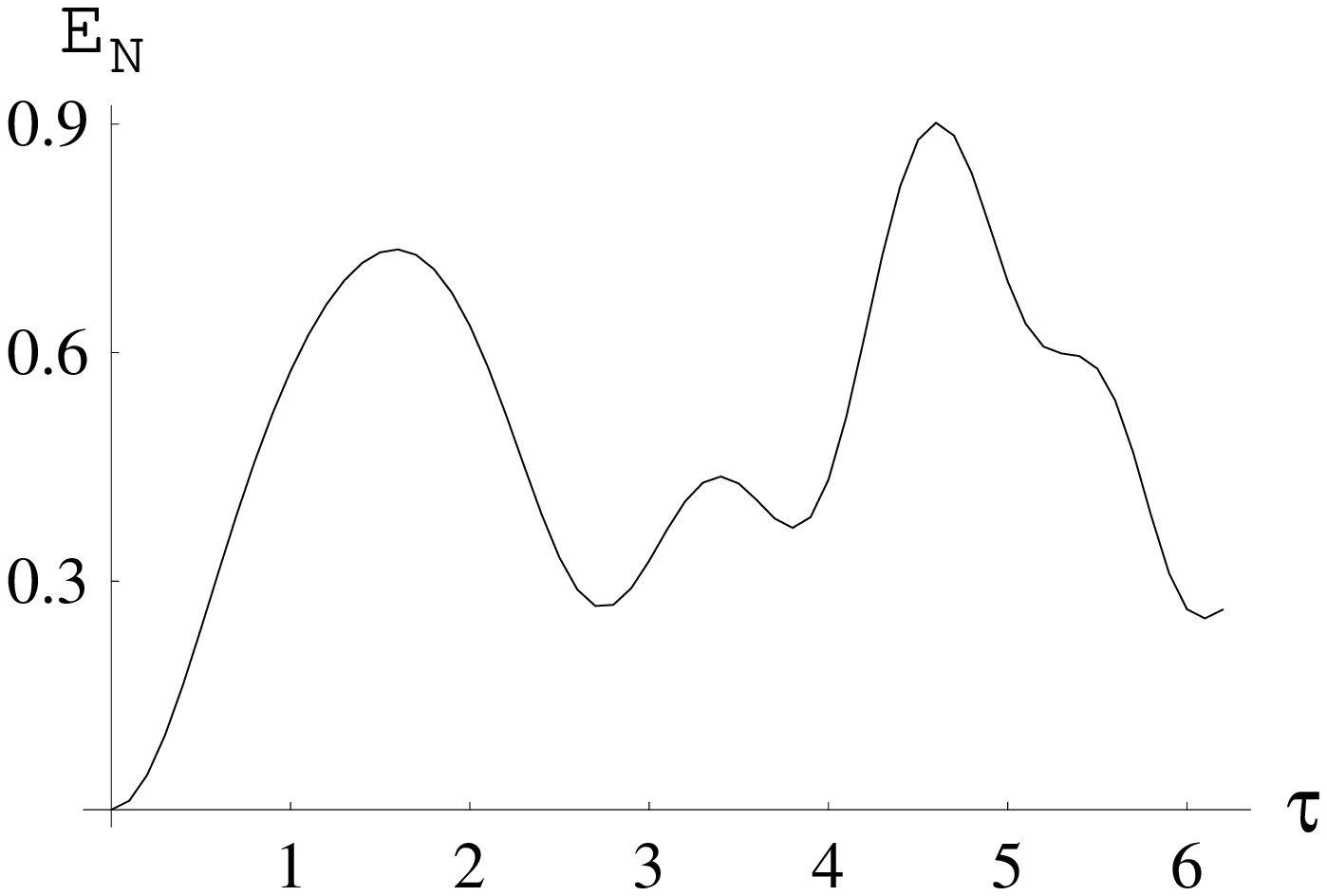,width=4.6cm,height=3.2cm}\psfig{figure=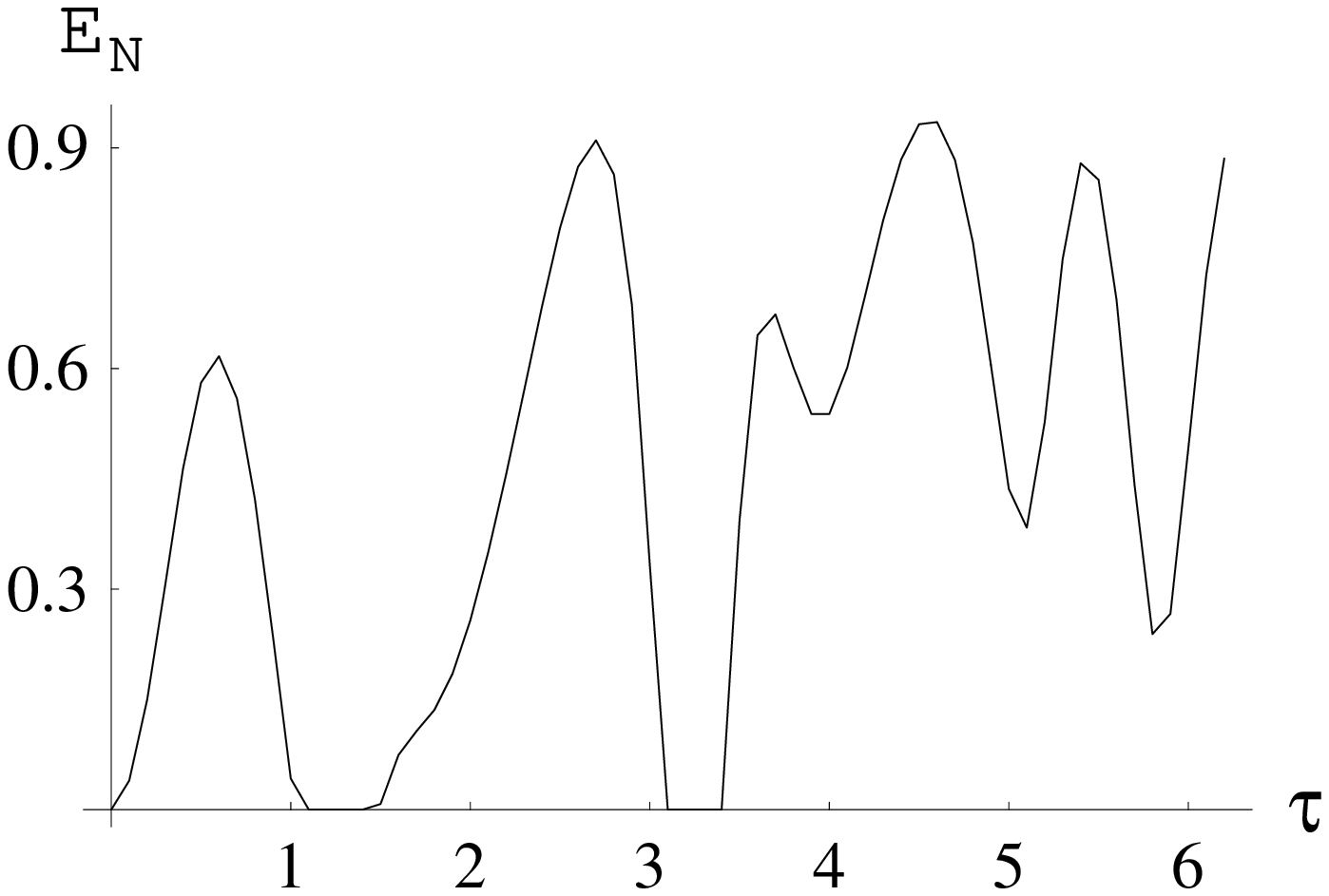,width=4.6cm,height=3.2cm}}
%\vspace*{-7cm}
\centerline{{(c)}\hskip4.0cm{(d)}}
%\vspace*{-2cm}
\centerline{\psfig{figure=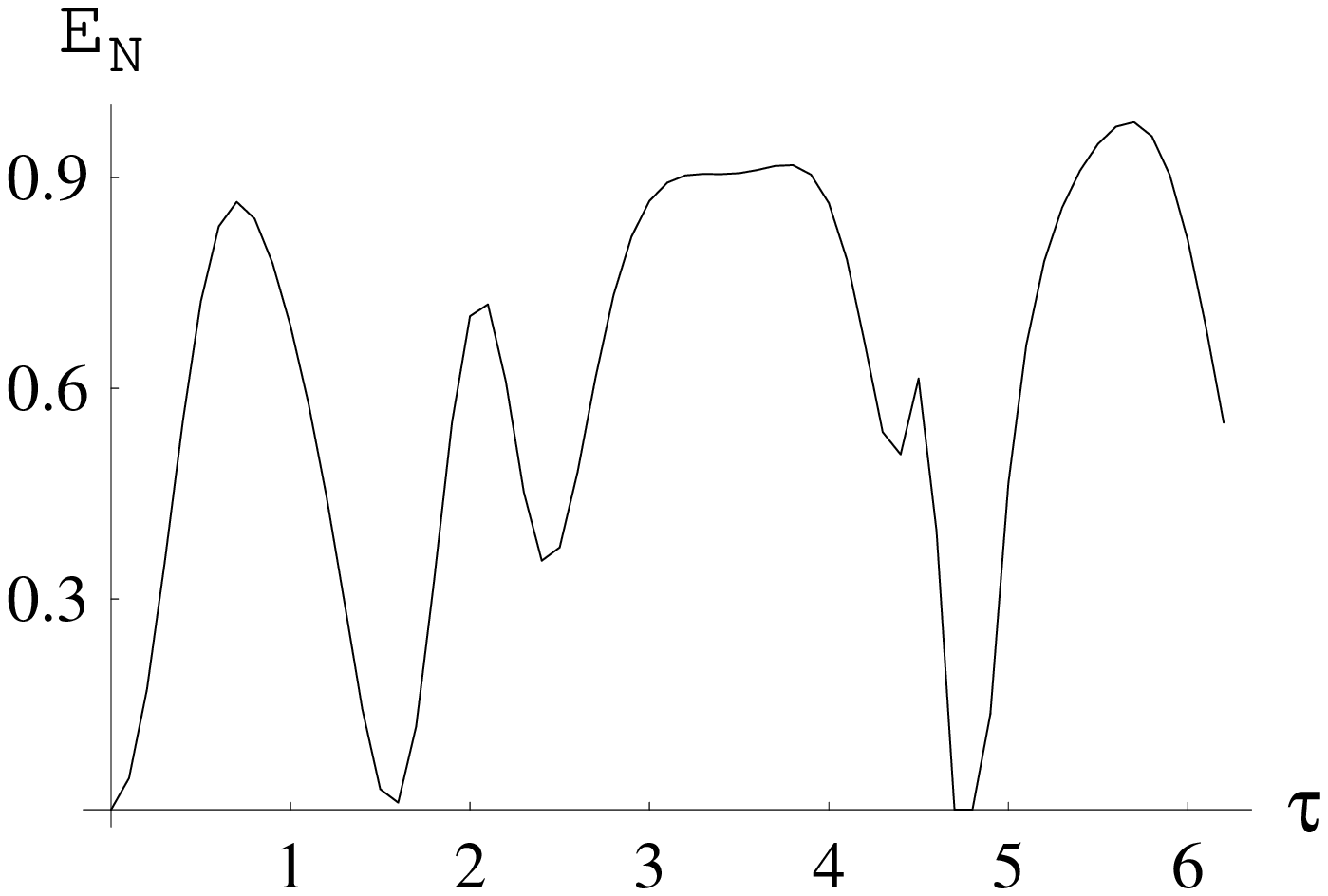,width=4.6cm,height=3.2cm}\psfig{figure=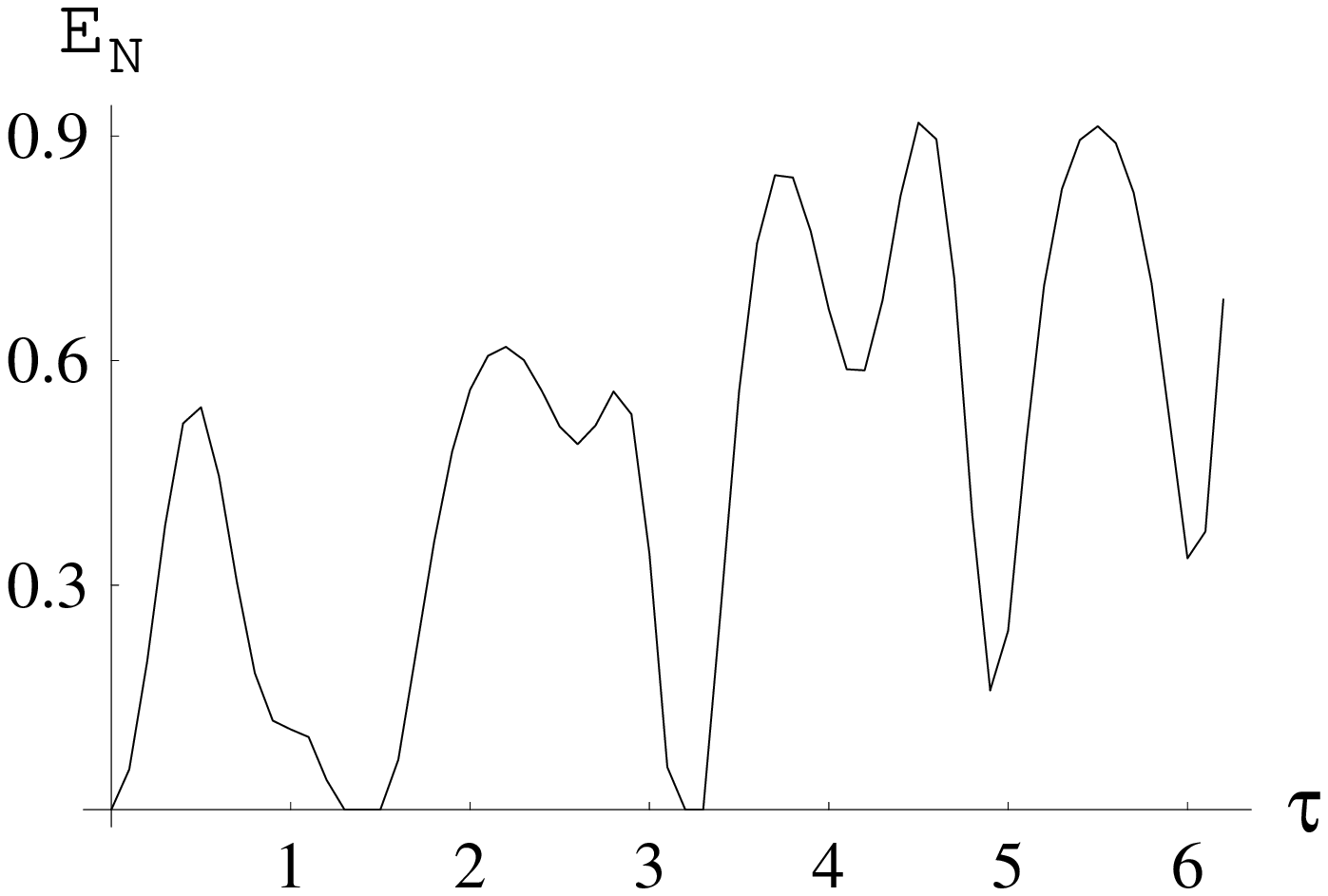,width=4.6cm,height=3.2cm}}
%\vskip-7cm
\caption{Entanglement against the rescaled interaction time $\tau_1=\tau$ for the case $\modul{\alpha}^2=0$ (with both the qubits $3$ and $4$ found in $\ket{1}$) and an initial squeezing parameter $r=0.86$. From {(a)} to {(d)} we show the behavior of the entanglement function for $\tau_2=0$, $\tau_2\simeq{\pi}$, $\tau_2\simeq{3\pi/2}$ and $\tau_2\simeq{2\pi}$.
All quantities plotted are dimensionless.} 
\label{risultato1}
\end{figure}

\begin{figure}[t]
%\centerline{{\bf (a)}\hskip4.0cm{\bf (b)}}
%\vskip-2cm
%\centerline{\psfig{figure=NUOVOsougator086a002diffuno.eps,width=7.6cm,height=12.2cm}\hspace*{-3cm}\psfig{figure=NUOVOsougator086a002diffdue.eps,width=7.6cm,height=12.2cm}}
%\vspace*{-7cm}
%\centerline{{\bf (c)}\hskip4.0cm{\bf (d)}}
%\vspace*{-2cm}
\centerline{\psfig{figure=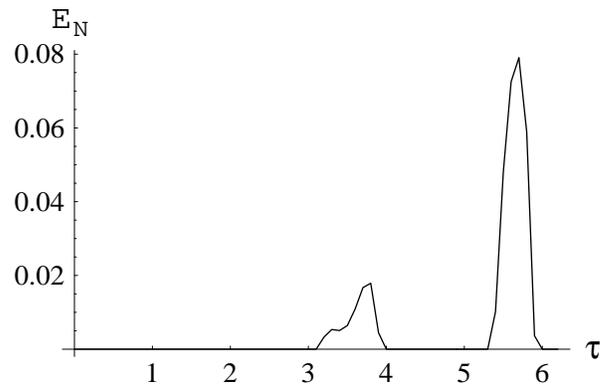,width=8.6cm,height=5.2cm}}
%\hspace*{-3cm}\psfig{figure=NUOVOsougator086a002diffquattro.eps,width=7.6cm,height=12.2cm}}
%\vskip-9cm
\caption{Positive part of the difference between the logaritmic negativity
for $\tau_1=\tau$ and $\tau_2=3\pi/2$ 
and the peak of logarithmic negativity
for $\tau_{1}=3\pi/2$ and $\tau_2=0$ (for which $E_{\N}\simeq{0.9}$) 
against the rescaled interaction time $\tau$. 
We have considered $r=0.86$ and $\modul{\alpha}^2=0$ as in the previous figure.
All quantities plotted are dimensionless.} 
\label{risultato2}
\end{figure}

In Fig.~\ref{risultato1}, we show the behavior of $E_{\cal N}$ between qubits $1$ and $2$ 
as a function of the rescaled interaction time $\tau_1=\tau$. The interaction time $\tau_2$ (relative to the coupling between qubits $3$ and $4$ 
and the CV system) is used as a parameter. We have considered $\alpha=0$ ({\em i.e.~}both the ancillary qubits are found in their excited state).  
As is apparent, the analytical structure of the transferred entanglement changes drastically in presence of the ancillary interaction and postselection 
which is, not surprisingly, not always advantageous for all interaction times as it can even completely spoil the entanglement between the qubits. 
However, large peaks of entanglement, with values beating  
the maximum obtained with the passive strategy (corresponding to $E_{\N}\simeq0.9$, see Table~\ref{table}), are frequently present.
In particular, as detailed in Fig.~\ref{risultato2}, for the interaction times $\tau_1=5.65$ and $\tau_2=3\pi/2$, 
the considered postselection process results in $E_{\cal N}\simeq 0.975$ with 
probability $p\simeq0.3$, thus allowing for the distillation of a {\it quasi-maximally entangled state}. 

\begin{figure}[t]
\centerline{{(a)}\hskip4.0cm{(b)}}
%\vskip-4cm
\centerline{\psfig{figure=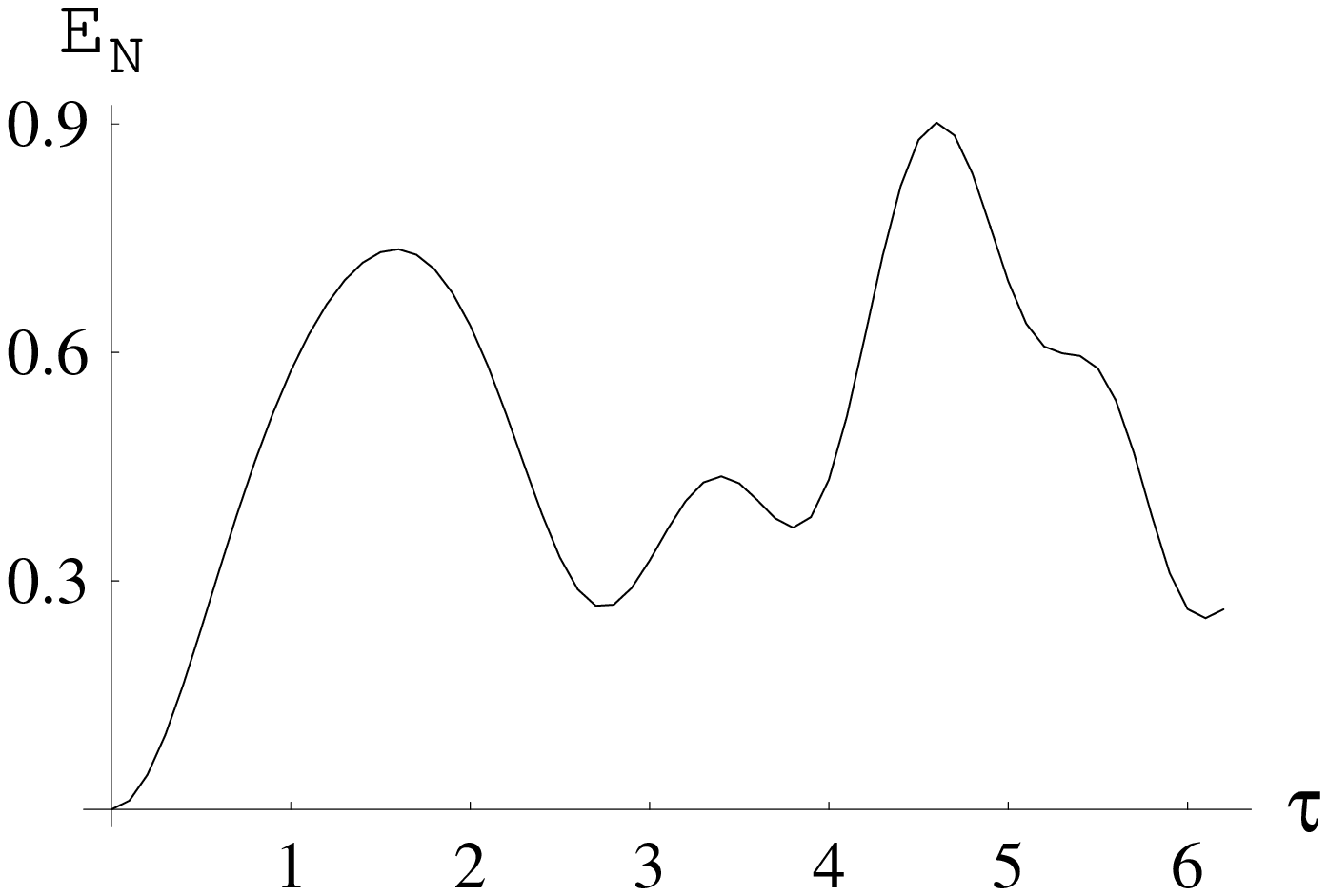,width=4.6cm,height=3.2cm}\psfig{figure=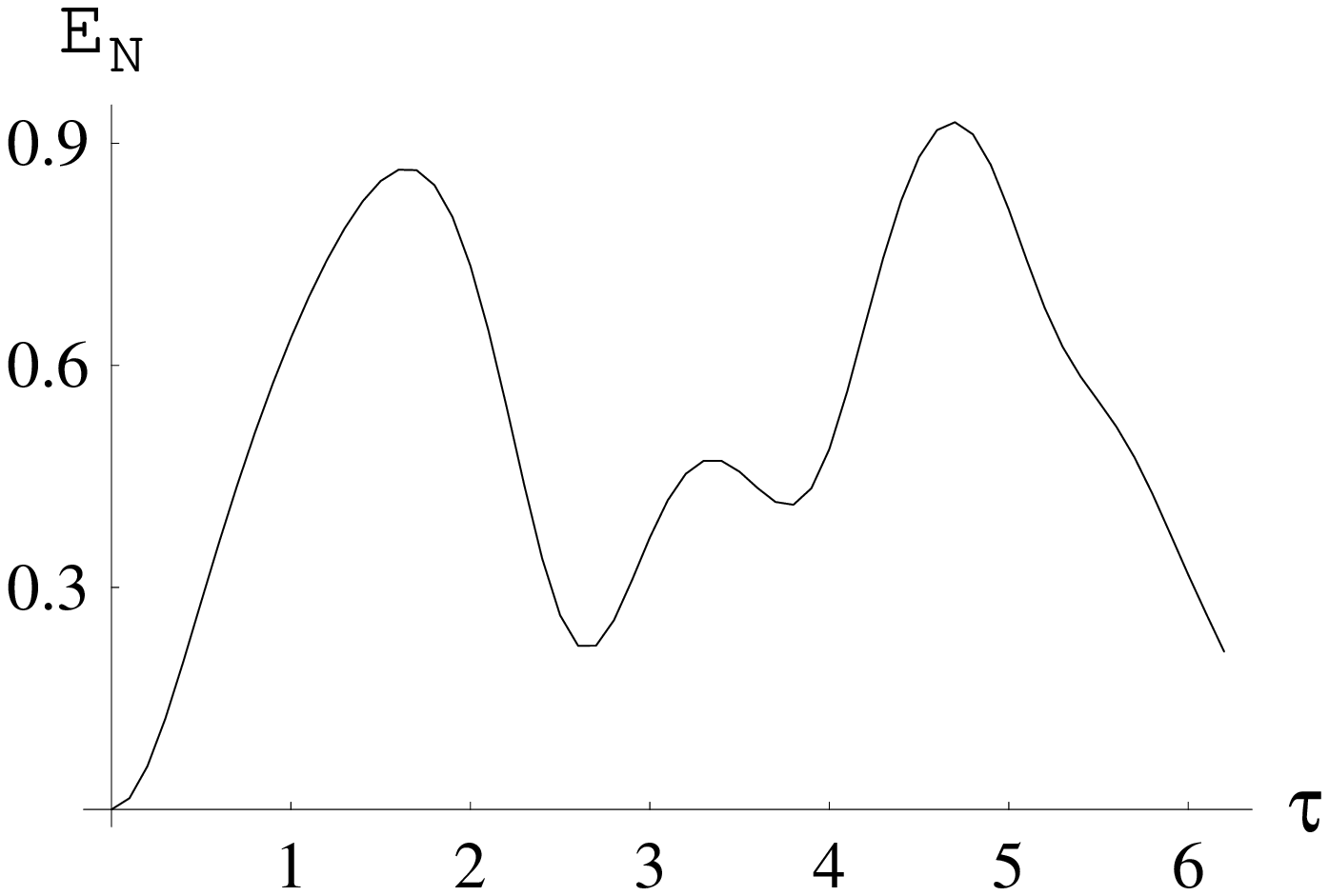,width=4.6cm,height=3.2cm}}
%\vspace*{-9cm}
\centerline{{(c)}\hskip4.0cm{(d)}}
%\vspace*{-4cm}
\centerline{\psfig{figure=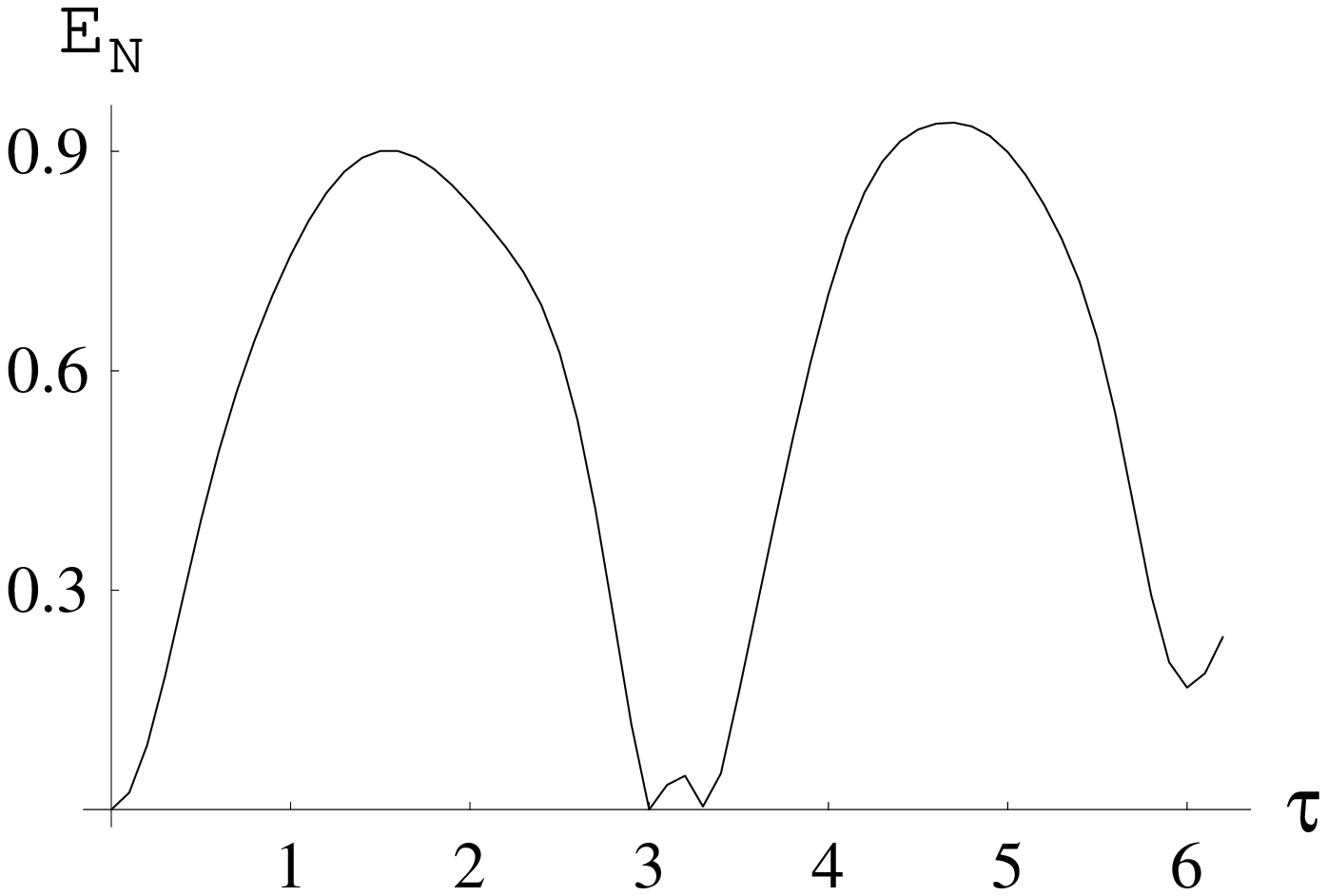,width=4.6cm,height=3.2cm}\psfig{figure=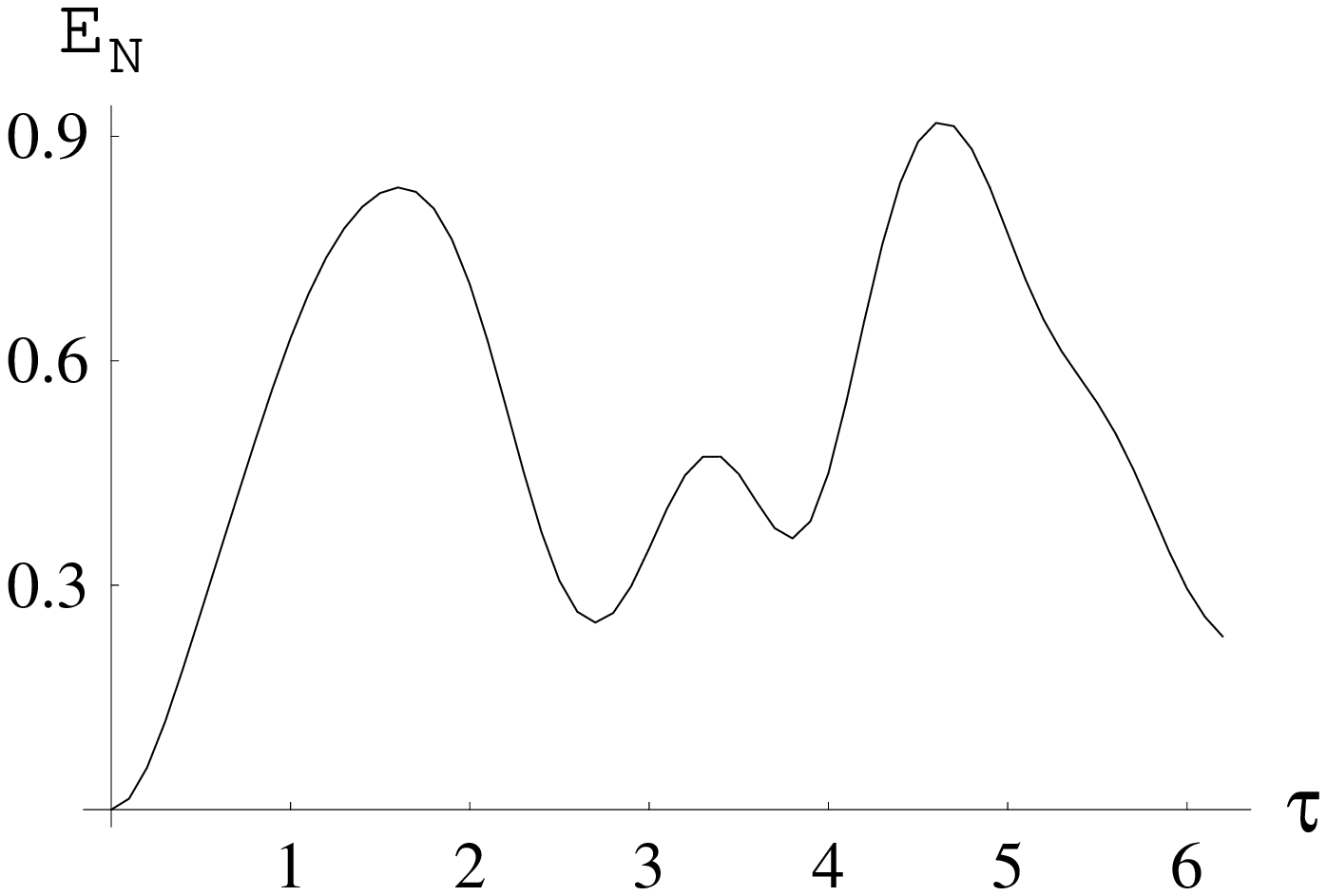,width=4.6cm,height=3.2cm}}
%\vskip-9cm
\caption{Entanglement against the rescaled interaction time $\tau_1=\tau$ for the case $\modul{\alpha}^2=0.95^2$ and an initial squeezing parameter $r=0.86$. From {(a)} to {(d)} we show the behavior of the entanglement function for $\tau_2=0$, $\tau_2\simeq{\pi}$, $\tau_2\simeq{3\pi/2}$ and $\tau_2\simeq{2\pi}$.
All quantities plotted are dimensionless.} 
\label{risultato3}
\end{figure}

Before addressing the refining of such almost perfect Bell pairs extraction procedure, 
some further remarks are in order. 
The procedure is qualitatively effective regardless of the choice of $\alpha$. Any value of $\alpha$ allows to beat the maximum of transferred entanglement achievable with the passive strategy (in Fig.~\ref{risultato3} we report the analysis for $\alpha=0.95$; in particular, panel ${(c)}$ corresponds to an increase in entanglement with respect to $0.9$ which can be as large as $3\%$ for $\tau_{1}=5.65$ and $\tau_{2}=3\pi/2$). 
However, as seen in panel (b), the changes in the behaviour of the 
transferred entanglement are much less abrupt  
(no qualitative changes with respect to the passive case of $\tau_1=0$ are found up to $\tau_{2}=1$)
and the values of the maximally transferred entanglement are lower 
than the corresponding ones obtained for $\alpha=0$. 
The optimality of the projection on the excited states of the ancillary qubits obtained for $\alpha=0$ can be understood at the level of the density matrix structure. Indeed, such a projection always results 
in the postselection of states of qubits $1$ and $2$ of the following form:
\begin{equation}
\label{desiderato}
\varrho_{cd,12}=
\begin{pmatrix}
A_1&0&0&-B\\
0&A_2&0&0\\
0&0&A_3&0\\
-B&0&0&A_4
\end{pmatrix}
\end{equation}
with the density matrix's entries being, in general, very complicated functions
of $\tau_{1,2}$ and $r$ (no qualitative information can be gathered, in general, from the presentation of their analytical form). Now, for optimal choices of $\tau_1$ an $\tau_2$, corresponding to the values discussed in Figs.~\ref{risultato1} (c) and~\ref{risultato2}, one has $A_{1},A_{4},B\gg A_{2,3}>0$ 
so that $\varrho_{cd,12}$ is a highly pure state with a 
very large projection on the Bell state $(\ket{00}-\ket{11})/\sqrt{2}$.
On the other hand, as soon as $|\alpha|>0$, contributions different 
from Eq.~(\ref{desiderato}) appear, becoming increasingly relevant for larger values of $|\alpha|$, thus reducing the efficiency of the entanglement transfer. 

The efficiency of the postselected protocol depends on the initial squeezing of the CV state
as well. 
Indeed, if 
we consider a small degree of initial entanglement between the field modes of the CV system, it can be analytically seen that the entanglement between 
qubits $1$ and $2$ in the postselected protocol described so far is not larger than the entanglement transferred through the passive protocol of Refs.~\cite{wonmin,ioMEMS,ioJJ}. In particular, by considering the expansion of Eq.~(\ref{2msq}) into terms up to ${\cal O}(r^2)$ and the joint qubits-CV modes dynamics, the postselected density matrix ($\alpha=1$ being in this instance the most favorable case) turns out to be equivalent to the non-postselected one. Choices of $\alpha$
different from $1$ correspond, in this case, to a smaller overlap between the reduced state of the qubits and $(\ket{00}-\ket{11})/\sqrt{2}$.

%\begin{figure}[t!]
%\psfig{figure=sougator086a095uno.eps,width=8.6cm,height=3.2cm}
%\psfig{figure=sougator086a095due.eps,width=8.6cm,height=3.2cm}
%\caption{Entanglement against the rescaled interaction time $\tau_1$ for the case of a small probability $\modul{\alpha}^2=0.95^2$ of finding both the qubits $3$ and $4$ in $\ket{0}$ and an initial squeezing parameter $r=0.86$. From {\bf (a)} to {\bf (d)} we show the behavior of the entanglement function for $\tau_2=0$, $\tau_2\simeq{\pi}$, $\tau_2\simeq{3\pi/2}$ and $\tau_2\simeq{2\pi}$.} 
%\label{risultato3}
%\end{figure}

\begin{figure}[t]
%\vspace*{-2cm}
\centerline{\psfig{figure=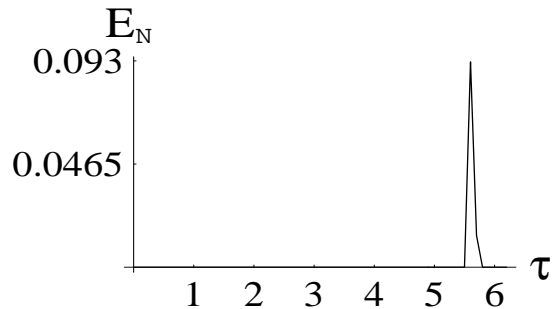,width=7.6cm,height=4.2cm}}
%\vspace*{-9cm}
\caption{Positive part of the difference between the logarithmic negativity 
for $\tau_{1}=\tau$ and $\tau_{2}\simeq{\tau_{3}}\simeq{3}$ and the passive peak value 
(for which $E_{\N}\simeq0.9$ against the interaction time $\tau$. 
An enhancement of about $0.093$ is found, 
which improves the result for a single iteration protocol shown in Fig.~\ref{risultato2}. 
We have considered again $r=0.86$ and $\modul{\alpha}^2=0$. All quantities plotted are dimensionless.}
\label{risultatoESTENSIONE}
\end{figure}

Let us now note that, as a matter of principle, the entanglement transfer
to a pair of two-level systems can be further (probabilistically) improved 
by considering the interaction of the entanglement distributor with 
additional pairs of ancillary qubits. To be more specific,
we have explicitly considered the case of a third pair of qubits, labelled with $5$ and $6$, 
which interact for a rescaled time $\tau_{3}$ with modes $a_1$ and $a_2$, after 
the measurement of the state of qubits $3$ and $4$ has been accomplished. 
We assume that the third pair of qubits are also 
found to be in $\ket{\psi_{\alpha}}\otimes\ket{\psi_{\alpha}}$. It is possible to choose $\tau_{2}$ and $\tau_{3}$ so that the maximum of the entanglement transferred by this active sequential protocol is even larger than what is shown in Fig.~\ref{risultato1} (c). 
Indeed, for $\tau_{1}\simeq5.65$, $\tau_{2}=\tau_3={{\pi}}$ (and $r=0.86$)
the transferred entanglement reaches  ${E}_{\cal N}\simeq0.99$, 
with a difference from the 
passive transferred entanglement enhanced to $0.093$ (see Fig.~\ref{risultatoESTENSIONE}).

This result is {\it per se} interesting as it permits to conjecture that the sequentialization 
of the postselected protocol would allow for an asymptotically perfect 
entanglement extraction. 
Moreover, quite remarkably, such an iterative process is efficient, in that 
a relatively small number of iterations does provide an almost perfect transfer
(as we have shown, the results are 
nearly perfect already after two iterations). Clearly, with respect 
to the first iteration of the postselection process, the second one is much less efficient, in that the entanglement enhancement is a much smaller 
fraction of the total transferred entanglement. 
This reduction in efficiency is a feature characteristic of any `distillation-like' (in a broad sense) process~\cite{ioNJP} and sets 
a practical trade-off between the amount of entanglement one would like to extract from the 
distributor and the number of iterations allowed by the decoherence time which characterizes 
the specific setup adopted. In many practical cases, like the one we are addressing, 
just two repetitions of the protocol could be enough to extract a nearly entire ebit.

%%%%%%%%%%%%%%%%%%%%%%%%%%%%%%%%%%%%%%%%%%%%%%%%%%%%%%%%%%
\section{Conclusions and outlook}
\label{finale}

We have considered the transfer and extraction of CV entanglement by Jaynes-Cumming interactions with 
finite dimensional atomic-like systems. Our results show that microscopic systems or, from a slightly 
different standpoint, systems with Hilbert spaces of `small' dimensions ($\lesssim20$) allow for an almost complete 
extraction of the CV entanglement originally stored in a infinite-dimensional Hilbert space. 
On the other hand, the reasons for the impossibility of a perfect extraction have been clearly 
shown to lie in the feature of the considered dynamics, which cannot perfectly mimic 
that of a quantum harmonic oscillator. In this respect, the macroscopic highly polarised limit 
(achievable in atomic clouds) can be already considered as optimal. However, it could be interesting 
to envisage a realistic microscopic system in which the couplings between different levels do actually 
reproduce the harmonic oscillator's ones, described by truncated bosonic ladder operators. 
Such a development could deserve further investigation. 

Furthermore, we have shown that a simple and realistic sequential postselection protocol 
would allow for the arbitrarily perfect extraction of the CV entanglement
into a Bell state and, even more significantly, that one iteration of such a protocol would 
provide with an almost perfect symmetric Bell state. Such a strategy could be a promising candidate to build up entangled pairs at a distance, with remarkable improvements over passive purely dynamical
strategies.

\acknowledgments

This work has been supported by The UK EPSRC and the Korea Research Foundation (Grant 2003-070-C00024). 
This research is part of QIP IRC www.qipirc.org (GR/S82176/01), through which
AS is supported. MP thanks The Leverhulme Trust (ECF/40157) for financial support. We thank M.E.~Reuter for his suggestions on the manuscript.

%%%%%%%%%%%%%%%%%%%%%%%%%%%%%%%%%%%%%%%%%%%%%%%%%%%%%%%%%%


\begin{thebibliography}{99}
\bibitem{proviso}Actually, the states measured in Ref.~\cite{laurat} are generally mixed and 
are not described by \eq{2msq}. In particular, the mixedness of the CV state would slightly hamper 
the entanglement transfer (as the thermal noise present implies an excitation of higher number states
and thus a possible mismatch of the dimensions of the effective supports of the two systems).
Anyway, considering a state with equal logarithmic negativity allows us to provide a good 
estimate of the number of atoms necessary to obtain a substantial extraction of the CV entanglement.
\bibitem{commento1} This assumption can be easily relaxed giving rise to an asymmetrical post-selection scheme. However we have found that, for the symmetric entangler state of Eq.~(\ref{2msq}), the best results in terms of entanglement transfer are found for the totally symmetric protocol of two identical measurements.
\end{thebibliography}
\end{document}